**Giant Photoluminescence Enhancement in MoSe$_2$ Monolayers treated with Oleic Acid Ligands**

Arelo O.A Tanoh[1,2], Jack Alexander-Webber[3], Ye Fan[3], Nicholas Gauriot[1], James Xiao[1],

Raj Pandya[1], Zhaojun Li[1], Stephan Hofmann[3], Akshay Rao[1]*

[1]Cavendish Laboratory, Cambridge, JJ Thomson Avenue, CB3 0HE, Cambridge, United Kingdom

[2]Cambridge Graphene Centre, University of Cambridge, 9 JJ Thomson Avenue, Cambridge, CB3 0FA, Cambridge, United Kingdom

[3]Department of Engineering, University of Cambridge, JJ Thomson Avenue, CB3 0FA Cambridge, United Kingdom

*E-mail: ar525@cam.ac.uk

**Abstract**

The inherently low photoluminescence (PL) yields in as prepared transition metal dichalcogenide (TMD) monolayers are broadly accepted to be the result of atomic vacancies (*i.e.* defects) and uncontrolled doping, which give rise to non-radiative exciton decay pathways. To date, a number of chemical passivation schemes have been successfully developed to improve PL in sulphur based TMDs *i.e.* molybdenum disulphide (MoS$_2$) and tungsten disulphide (WS$_2$) monolayers. Studies on solution based chemical passivation schemes for improving PL yields in selenium (Se) based TMDs are however lacking in comparison. Here, we demonstrate that treatment with oleic acid (OA) provides a simple wet chemical passivation method for monolayer MoSe$_2$, enhancing PL yield by an average of 58 fold, while also improving spectral uniformity across the material and reducing emission linewidth. Excitation intensity dependent PL reveals trap-free PL dynamics dominated by neutral exciton recombination. Time-resolved PL (TRPL) studies reveal significantly increased PL lifetimes, with pump intensity dependent TRPL measurements also confirming trap free PL dynamics in OA treated MoSe$_2$. Field effect transistors show reduced charge trap density and improved on-off ratios after treatment with OA. These results indicate defect passivation by OA, which we hypothesise act as ligands, passivating chalcogen defects through oleate coordination to Mo dangling bonds.

Two-dimensional (2D) (*or* monolayer) transition metal dichalcogenides (TMDs) continue to attract wide-spread research interest due to their intriguing optical and electronic properties.[1–3] Few to hundreds of microns sized monolayers can be isolated from their layered bulk counterparts by overcoming the interlayer van der Waals interaction *via* various layer by layer exfoliation methods. These include dry mechanical cleavage;[3,4] metal assisted exfoliation[5] and; liquid phase exfoliation (LPE).[6] There are also continuous efforts to grow wafer-scale crystalline monolayers *via* epitaxial growth methods such as chemical vapour deposition (CVD).[7] A number of TMDs transition from indirect optical gap in the bulk crystal to direct optical gap as a monolayer.[8] The direct optical gap, high absorption[9] and potentially high charge carrier mobilities of a number of monolayer TMDs has spurred research into their application to optoelectronic devices namely photodetectors, light emitting diodes (LEDs),[10] field effect transistors (FETs)[2] and on-chip single photon quantum emitters.[11] Moreover, the massively reduced dielectric screening gives rise to tightly bound excitons[1,8] even at room temperature, thus providing a convenient means to study the many body exciton-exciton and exciton-charge interaction that give rise to a multitude of exotic neutral excitons[12] and charged excitons.[13,14]

Although monolayer TMDs hold great promise for future optoelectronic applications, as-prepared monolayers tend to exhibit low photoluminescence quantum efficiency (PLQE).[3,10] The persistence of non-radiative pathways in pristine monolayers has been attributed to chalcogen (*i.e.* S and Se) vacancies,[15,16] atomic substitutions[17,18] and trion formation.[15,19] Methods to improve material performance broadly take two routes; encapsulation or chemical passivation. Encapsulation exploits the atomically flat dielectric properties of hexagonal boron nitride (hBN), using it as an encapsulation medium[20,21] or sub-layer[22] that isolates TMD monolayers from doping and disorder imposed by common substrate materials. This preserves their intrinsic properties and improves overall optical quality as given by spatially homogenous narrow linewidths in PL spectra. Encapsulation with hBN has been shown to suppress exciton-exciton annihilation in monolayer tungsten disulphide ($WS_2$), improving PL, however at high excitation intensities.[23] Large PL enhancement at low excitation density has not been demonstrated with hBN encapsulation alone.

On the other hand, recently, a number of successful chemical passivation schemes have been devised to enhance the PLQE of sulphur based TMDs, namely molybdenum disulphide ($MoS_2$) and tungsten disulphide ($WS_2$). Such methods include the use of p-doping agents such as 2,3,5,6-tetrafluoro 7,7,8,8-tetracyanoquinodimethane (F4TCNQ),[24,25] hydrogen peroxide,[26] or deposition of a monolayer titanyl phthalocyanine (TiOPc) charge transfer interface.[27] These techniques aim to withdraw electrons to suppress the formation of low PLQE trions, promoting dominant neutral exciton recombination. One of the most successful of these chemical treatments has been the use of the non-oxidizing `super acid' bis(trifluoromethane)sulfonimide (TFSI)[15,28,29] to treat $MoS_2$ and $WS_2$, leading to large increases in PL. It has been suggested that TFSI acts as a strong electron withdrawing (p-doping) species *via* comparative studies with gated n-type $MoS_2$ and $WS_2$ monolayers, whereby applying a negative bias suppresses non-radiative pathways *via* trion formation, leaving dominant neutral exciton recombination and similar PL dynamics to TFSI treated monolayers.[19] However, PL dynamics in TFSI treated $MoS_2$ and $WS_2$ have been shown to be trap-limited,[30,31] and it has recently been shown that this is due to presence of sulphur vacancies which remain unpassivated even with the TFSI treatment.[32]

Recently, the authors of this study demonstrated 26 and 20 fold increase in $WS_2$ PL and electronic mobilities respectively *via* surface treatment with Oleic Acid (OA) ligands, outperforming treatment with TFSI.[31] The OA treatment results in high spectral uniformity with non-trap limited PL dynamics compared with TFSI treated monolayers, which indicate defect passivation by OA ligands. In support of this, electrically gated monolayers treated with OA show increased field effect mobilities with reduced charge trap density and no additional doping in comparison to their untreated or `pristine' form. The study also revealed bright trion PL evolution in OA treated $WS_2$ at high excitation densities due to binding between untrapped excitons and local *n*-type charges. This strong trion evolution has potential applications in quantum information processing. The authors suggested defect passivation *via* dative covalent bonding between the oleate group on the OA ligand and metal atom at the chalcogen vacancy, which prevents defect/ trap assisted non-radiative exciton decay and promotes direct band-edge recombination- thus improving PL yields in a manner akin to defect passivation by OA in colloidal nanocrystals.

In contrast to the range of chemical treatments for sulphur based TMDs, there has been little success in developing treatments for selenium based TMDs *i.e.* molybdenum diselenide ($MoSe_2$) and tungsten diselenide ($WSe_2$).[28] For instance, TFSI is also known to quench PL in both these materials instead of enhancing it.[28] Han *et al.*[33] however achieved 30-fold enhancement of defect rich CVD $MoSe_2$ PL at room temperature *via* exposure to hydrobromic acid (HBr) vapour. The authors attribute this outcome to *p*-doping by the HBr combined with structural repair of chalcogen vacancies. Structural repair occurs *via* the replacement of oxygen substitutions by bromine (Br) ions at selenium (Se) vacancies which acts to suppress trapped exciton states, thus eliminating non-radiative pathways. Recently, high PLQE of as-prepared CVD $WSe_2$ has been demonstrated *via* a solvent evaporation-mediated decoupling (SEMD),[34] whereby the solvent evaporation process assists in the separation of as-grown synthetic monolayers from the underlying substrate. This serves as alternative to polymer assisted transfer methods, which involve the use of harsh chemicals e.g. hydrofluoric acid (HF). The drastic improvement in optical quality compared to standard CVD monolayer transfer techniques is considered to be related to overcoming substrate induced mechanical strain, which can introduce band structure modifications that reduce PL.[34,35] These methods however, do not provide the ease of processing that simple solution based chemical approaches do and rely on specific growth conditions, restricting their general purpose application.

Here, we demonstrate that oleic acid (OA) treatment of $MoSe_2$ monolayers results in greatly enhanced neutral exciton PL, as well as trap-free PL dynamics. In addition, OA treated $MoSe_2$ field effect transistors (FETs) exhibit marked improvement in transfer characteristics. The reduced subthreshold swing (SS) indicating reduced charge trap density and hence improved current on/off ratios. These results highlight the versatility of OA treatment and provides a simple solution based chemical passivation protocol for selenide based TMDs.

## Results & Discussion

Monolayer MoSe$_2$ samples were isolated from their bulk crystal *via* gold mediated exfoliation[5] onto Si-SiO$_2$ (90 nm SiO$_2$) for PL maps and FET characterization and; thin (~170 µm) glass cover slides for excitation intensity dependent PL, TRPL and Raman spectroscopy. All measurements were performed at room temperature. Following initial optical and electronic characterization, the untreated or `pristine' samples were coated with OA *via* drop casting in a nitrogen environment and left for 12 hours on a hot plate set at 25 °C. The treated samples were subsequently rinsed with anhydrous toluene and blow dried with a nitrogen (N$_2$) gun before further optical and electronic characterisation.

Figure 1.a. shows a cartoon illustration of OA ligand coordination to a selenium (Se) vacancy. Figure 1.b shows the scatter plot of spectral position of peak emission and PL integrals extracted from PL maps of multiple MoSe$_2$ monolayers on Si-SiO$_2$ substrates before and after OA treatment. Maps were measured at 126 W cm$^{-2}$. Figure 1.c shows the PL spectra for points on an exemplary monolayer that correspond to the median PL enhancement, $Δ_{median}$, where $Δ = PL_{after\ treatment}/ PL_{before\ treatment}$. Table 1 shows statistical information derived from figure 1.b, namely: average PL enhancement across the monolayers ($Δ_{ave}$); standard deviation in PL integral ($σ_{PL}$); average emission peak wavelength ($λ_{ave}$); and standard deviation in peak wavelength ($σ_λ$). The untreated case is indicated by (*). Figure 1.d. shows the normalised spectra from Figure 1.c, with single Gaussian peaks fitted to estimate the change in spectral linewidth between untreated (blue) and treated (red) cases.

**Table 1: PL enhancement statistics derived from PL maps of MoSe$_2$ monolayers. Characteristics prior to treatment marked with (*)**.

| $Δ_{ave}$ | $σ_{PL}$ | $λ_{ave}$ | $σ_λ$ |
|---|---|---|---|
| **58** | 56%* → 29% | 794 nm* → 787 nm | 3.31 nm* → 1.02 nm |

An average PL enhancement of 58 fold is observed upon OA treatment. The standard deviation in PL intensity decreases from 56% to 29%. This demonstrates that the OA treatment both improves PL and spatial homogeneity in brightness. A spectral narrowing is also observed with an average blue shift $λ_{ave}$ of 7 nm with improved spectral uniformity given by a 69% reduction in $σ_λ$ from the untreated to the treated case. The median PL enhancement was calculated as $Δ_{median}$ ~ 61 fold. The normalised median spectra show a blue shift in spectral peak of 12 nm (798 nm → 786 nm) and reduction in full width half maximum (FWHM) of 5.5 nm (27.2 nm → 21.7 nm) from the untreated to treated case. As also observed in OA treated WS$_2$,[31] the spectral blue-shift and line-width narrowing of MoSe$_2$ PL may be attributed to changes in strain induced by ligand coordination. These results however establish of the efficacy of the OA treatment to enhance the PL properties of MoSe$_2$.

To probe the exciton dynamics that accompany the PL enhancement, we look at the excitation intensity dependent room temperature PL of a monolayer before and after OA treatment. Figure 2.a-d shows the results derived from a room temperature steady state excitation intensity dependent PL series over five orders of magnitude. Intensities range between 0.018 W cm$^{-2}$ and 909 W cm$^{-2}$, staying well below 9000 W cm$^{-2}$ to avoid thermal damage.[15]

Figure 2.a. shows no noticeable changes in spectral properties with increased excitation intensity in the untreated monolayer. When treated with OA, Figure 2.b shows overall spectral narrowing compared to the untreated case, however no additional spectral components are observed, unlike the case of OA treated WS$_2$ which shows strong trion contribution at high excitation intensities.[31]

Figure 2.c shows a log-log plot of PL integral as a function of excitation intensity for untreated (blue) and treated (red) samples. The gradients (*m*) of the series represent the exponent to the power law fit, I = P$^m$.[28] As such the *m* values indicate the exciton recombination regimes observed. Figure 2.d. shows the ratio of PL to excitation intensity (*γ*), which serves as a relative PLQE value. At low intensities, the untreated sample shows slight super-linear behaviour ($m_1$ ~ 1.05), which is indicative of some degree of exciton trapping[31] between 0.06 W cm$^{-2}$ and 0.8 W cm$^{-2}$. This suggests a lack of non-radiative exciton-exciton annihilation, as given by the little variation γ ratio values between 1 W cm$^{-2}$ and 10 W cm$^{-2}$, albeit with low PLQE. Beyond 10 W cm$^{-2}$, the trend becomes sublinear ($m_2$ ~ 0.76), indicating the onset of non-radiative exciton-exciton annihilation.[15,19,28,29] However non-radiative trap assisted recombination processes dominate throughout the series, given the low PLQE of untreated TMD monolayers.[15,16,19,24–29,31]

When treated with OA, the emission follows a sub-linear power law exponent of *$m_1$ ~ 0.89* even at lower powers, signifying the immediate onset of non-radiative exciton-exciton annihilation and becomes more drastic at higher excitation intensities where *$m_2$ ~ 0.73*. These trends are reflected in the *γ* ratio which shows a general gradual reduction between 0.02 W cm$^{-2}$ and 0.76 W cm$^{-2}$ before sharply decreasing thereafter due to intensified exciton-exciton annihilation. The immediate exciton-exciton annihilation seen in the OA treated sample is consistent with trap-free exciton diffusion, similar to what has been observed with OA treated WS$_2$.[31] The increase in relative PLQE, γ, by an average factor of ~17 between 0.02 and 0.1 W cm$^{-2}$ also confirms significant reduction in non-radiative recombination *via* trap states.

We attempt to characterize the exciton species that contribute to PL of OA treated MoSe$_2$. Figure 3 a-c shows the results obtained from deconvoluting each PL spectrum in the OA treated MoSe$_2$ excitation intensity series (Figure 2.b). As per previous studies[31,36,37] Gaussian fits were used to identify the emissive excitonic species in the spectra. Figure 3 a shows the raw PL spectrum (red) taken in the high intensity regime (455 W cm$^{-2}$), where trion emission has been observed in OA treated WS$_2$ monolayers.[31] The dashed maroon and pink Gaussian fits represent the neutral exciton (*X*) and a low energy species (*ζ*) respectively. Figure 3.b. shows the excitation intensity series of *X* and *ζ*. It is clear that assigned neutral exciton (*X*) and lower energy species (*ζ*) obey the same recombination dynamics, with the same power law exponents (*m*) seen in Figure 2.c. Figure 3.c shows the ratio of *ζ* to *X* as a function of excitation intensity, which remains fairly constant and < 1 throughout the series.

The constant *ζ/X* < 1 ratio indicates the dominance of neutral excitons compared to low energy species such as trions[38] throughout the series. Trions in particular, evolve from the binding of neutral excitons with free photoionized charges and have been characterized in room temperature WS$_2$ PL measurements, which show the growth of a broadening and red-shifting low energy feature as a function of increasing excitation intensity.[31,36–38] While strong neutral exciton contributions are observed throughout the series, easily discernible trion evolution is not apparent in both pristine and OA treated MoSe$_2$ PL spectra. A recent study on exciton and trion dynamics in MoSe$_2$ concluded that trion formation is suppressed at room temperature due to changes in localisation effects.[39] To this end, OA treatment simply improves neutral exciton PL by reducing the density of non-radiative channels which may take the form of trap states caused by chalcogen vacancies. As per the work cited,[39] identifying the effects of OA treatment on trion emission on MoSe$_2$ would require low temperature PL studies.

To gain further insights into the exciton dynamics present in OA treated $MoSe_2$ we employ time-resolved photoluminescence (TRPL) microscopy. Figure 4.a shows normalized PL decay signals at room temperature under comparable low intensity 550 nm, 5 MHz pulsed laser excitation. Signal was detected using a visible rage single photon avalanche diode (VIS-SPAD) *via* 750 nm long pass and 900 nm short pass filters, removing laser excitation and collecting $MoSe_2$ PL only. Pulsed excitation intensities used were 0.054 W cm$^{-2}$ and 0.064 W cm$^{-2}$ for pristine (blue) and OA treated (red) cases respectively. Both signals are best described by a bi-exponential decay model (black dashed lines) consisting of a fast $\tau_1$ and slow $\tau_2$ components. The pristine sample (blue) PL decays with $\tau_1$ ~ 1.07 ns and $\tau_2$ ~ 3.06 ns. For the OA treated case (red), PL lifetimes are extended by a factor of 3x and 3.8x vs the pristine sample for fast and slow decays respectively with $\tau_1$ ~ 3.3 ns and $\tau_2$ ~ 12.07 ns. The overall increase in PL lifetimes due to OA treatment vs the pristine case reveals a suppression of non-radiative decay channels.

Figure 4.b. shows the variation in fast decay component $\tau_1$ as a function of initial carrier concentration $n_0$ over four orders of magnitude. Initial carrier concentrations were computed using openly available $MoSe_2$ steady state absorption data.[28] The excitation intensities used fall within the range used for the steady state excitation series shown in Figure 2 c-d. As shown in supplementary information (SI) Figure 2, all decays in the series fit a bi-exponential model. The pristine case shows very little variation in $\tau_1$ over the range of $n_0$ which indicates exciton trapping. In contrast, OA treated $MoSe_2$ shows a general reduction in $\tau_1$ for in the $n_0$ range measured. This lies in agreement with the sub-linear trend measured within the same excitation intensity regime shown in Figure 2.c, which indicates the immediate onset of exciton-exciton annihilation at low excitation fluences. Accordingly, the observed reduction in $\tau_1$ as a function of $n_0$ in the OA treated sample implies non-trap limited movement of excitons and thus provides further evidence for trap state passivation due to OA treatment. SI figure 3 shows the equivalent comparison for slow decay component, $\tau_2$, where the pristine case shows a general increase in lifetime as a function of $n_0$ in accordance with trap state filling while the OA treated sample shows a reduction in lifetime as a function of $n_0$.

In summary, steady state PL measurements presented so far show that OA treatment greatly enhances the PL of monolayer $MoSe_2$ and optical quality in terms of emission linewidth and spatial homogeneity in brightness. Steady state excitation intensity dependent PL and TRPL studies reveal trap-free neutral exciton movement in OA treated $MoSe_2$. The observed enhanced PL and trap free exciton annihilation dynamics combined support the hypothesis of true defect passivation by OA.

The exact surface chemistry that gives rise to the observed optical improvement is not fully clear at the moment and future experimental and theoretical studies will be required to understand the underlying mechanism. We note that Raman spectra of pristine and OA treated $MoSe_2$ in SI Figure 4 show no distinguishable structural changes. We however consider that the treatment mechanism would be linked to passivation of chalcogen defects through oleate coordination to Mo dangling bonds. Chemical passivation of these vacancy sites suppresses excitonic trap states, resulting in vastly improved PL efficiency due to direct band-edge recombination. In addition, formation of an OA layer with bulky alkyl chains may provide an insulating encapsulant of the TMD monolayer analogous to hBN encapsulation, resulting in better protection from reactive species formed from atmospheric oxygen and water, and external trap states outside of the monolayer from adsorbants.

Finally, to assess the impact of OA treatment on the electronic properties of monolayer MoSe$_2$, we test back-gated field effect transistors. Figure 5.a shows the predominantly *n*-type transfer characteristics of MoSe$_2$ before OA passivation, consistent with previous reports.[40] The as-fabricated devices are then treated with OA under the same conditions as described in the experimental section. After OA treatment *n*-type transfer characteristics are preserved. There is a relatively small threshold voltage ($V_{th}$) shift from $V_{th,Un}$ = 4.8 ± 1 V to $V_{th,OA}$ = 1.3 ± 2.3 V (Figure 5.b) indicating no substantial change in doping induced by the OA treatment. After OA treatment, devices consistently show an improved subthreshold swing (*SS*) from $SS_{Un}$ = 4 ± 0.9 V/dec to $SS_{OA}$ = 1 ± 0.1 V/dec (Figure 5.c), which indicates a reduction in interface charge trap density and is consistent with the notion of defect passivation by OA. A higher on-state current, due to reduced charge trapping, and larger off-state resistance after OA treatment leads to an improved on-off current ratio up to ~$5\times10^4$ (Figure 5.d).

## **Conclusions**

In conclusion, we have established OA surface treatment of MoSe$_2$ as an effective means of achieving drastically improved PL yield and trap free PL dynamics, as compared with untreated monolayers. PL statistics reveal that OA treatment yields monolayers of improved optical quality by way of bright spatially homogenous PL with narrow spectral linewidth. A steady state excitation intensity dependent PL series reveals significantly improved `PLQE' with trap-free exciton dynamics, which is taken as initial evidence of passivation of non-radiative trap states by OA ligands. Analysis of the excitonic species present in the excitation intensity series verifies dominant neutral exciton recombination in OA treated MoSe$_2$ under low to high excitation intensities. Consistent with improved steady state PL, time resolved PL studies reveal significantly improved PL lifetimes. The reduction in PL lifetimes as a function of initial carrier concentration also indicates trap free exciton movement, which further supports the hypothesis of PL enhancement as a result of ligand passivation. By way of surface chemical interaction between OA and monolayer MoSe$_2$, we hypothesise that the OA ligands coordinate to Mo dangling bonds at Se vacancies, which are known to be exciton trap states, and thus passivating them and yielding increased radiative efficiency. The insulating ligands may also protect the monolayer from atmosphere induced doping and surface induced strain, thus acting as an encapsulant, which may also contribute the PL linewidth narrowing. OA treated MoSe$_2$ based FETs show no significant additional doping. However, we observe a considerable improvement in subthreshold swing with orders of magnitude increase in on-off ratio, which provides further evidence of trap or defect passivation by OA. In essence, the results show that OA treatment is an effective, simple and versatile `wet' chemistry technique than can improve the PL characteristics of a selenide based TMD. Combined with previous studies on sulphur based TMDs, these results establish the 'ligand' based passivation approach as a universal defect treatment protocol for both sulphide and selenide based TMDs.

**Methods**

Sample Preparation

Prior to exfoliation, substrates were solvent processed *via* sonication in acetone and isopropyl alcohol (IPA) for ~15 mins and treated in $O_2$ plasma to remove adsorbants. Silicon-silicon dioxide (Si-SiO$_2$) substrates with 90 nm oxide layer were used for steady state PL, while thin (~ 170 μm) 22 mm x 22 mm glass cover slides were used for steady state excitation intensity dependent PL, TRPL and Raman microscopy.

Large area MoSe$_2$ monolayers were prepared *via* gold-mediated exfoliation.[5] The bulk crystal purchased from 2D Semiconductors was exfoliated manually onto low density clean-room tape prior to depositing a thin gold layer (~100-150 nm) *via* thermal evaporation under vacuum conditions. Following gold evaporation, thermal release tape was adhered to the gold coated MoSe$_2$ exfoliate, whereupon the cleanroom tape was peeled off, leaving the top-most layer of MoSe$_2$ attached to the gold on thermal tape. The thermal tape was then stuck onto the freshly plasma treated target substrate and heated on a hot plate up to 125 °C, pealing the thermal tape and leaving the TMD monolayers sandwiched between the gold and substrate. The excess gold was removed by immersing in Potassium Iodide (KI$_2$) and Iodine (I$_2$) standard gold etch (Sigma Aldrich). The substrate was gently swirled in the etchant for 5 minutes prior to rinsing in deionised water, 10-minute sonication in acetone and 5-minute rinse in isopropyl alcohol (IPA). Samples were dried with a nitrogen (N$_2$) gun. Monolayers were initially identified *via* the optical contrast method [41] and PL microscopy.

OA treatment

In a nitrogen glovebox, degassed OA (sigma Aldrich) was drop casted onto the exfoliated monolayer TMD samples and placed on a hot plate set to 25 °C for 12 hours overnight. Following treatment, the samples were rinsed in anhydrous toluene and blow dried with a N$_2$ gun.

Steady state PL microscopy

PL spectroscopy was performed on a Renishaw Invia confocal setup equipped with motorized piezo stage, using an air-cooled Ar-ion (Argon ion) 514.5 nm continuous wave (CW) laser *via* 50× objective (NA = 0.75). Signals were collected in reflection *via* notch filter. The diffraction limited beam spot size was estimated as 0.84 μm. PL signal was dispersed *via* a 600 l/mm grating prior to detection with inbuilt CCD detector. Laser power was measured directly *via* 5× objective with a Thorlabs S130C photodiode and PM100D power meter.

PL maps were generated from multiple MoSe$_2$ monolayers before and after OA treatment. Maps were generated with 1 μm resolution and 0.5 s integration time at 0.7 μW. Steady state intensity dependent PL series were performed on the MoSe$_2$ monolayers on glass slide. Care was taken to measure PL in the same locations before and after treatment. The spatial x, y and z focal plane of the measurement site and photodiode were recorded to accurately switch between photodiode and monolayer location for each PL measurement using the WIRE software piezo stage control interface. PL signals were scaled up to 500 s integration time as used on the lowest excitation intensity. Dark counts were recorded with the same integration times used for each PL measurement in the series. Dark counts were scaled and subtracted from the raw PL data accordingly.

A single Gaussian fit from the standard peak fit library in Origin lab was used in Figure 1.d. to estimate changes in FWHM between untreated and OA treated MoSe$_2$ PL signals. Exciton species in Figure 3.a. were deconvoluted from OA treated MoSe$_2$ PL signals with dark counts subtracted using a procedure written in Matlab, which incorporates the `gauss2' two Gaussian model fit. Further information on the Gaussian model is available on the *mathsworks* website.

Raman microscopy

MoSe$_2$ monolayers on thin glass cover slides were characterized *via* Raman spectroscopy using a Renishaw Invia confocal setup. Excitation was provided using a 530 nm CW laser *via* 50× objective (NA = 0.75), producing an estimated diffraction limited beam spot size of 0.86 $\mu$m. The Raman signal was collected in reflection *via* notch filter and dispersed with a 1800 l/mm grating prior to detection with inbuilt CCD camera.

Time resolved PL microscopy

TRPL measurements were performed using a custom built inverted PL microscope setup equipped with a motorized piezo stage. Excitation was provided by a pulsed super continuum white light source (Fianium Whitelase) filtered *via* a Bentham TMc 300 monochromator. TRPL excitation was acquired using a 550 nm laser *via* 60x oil objective, producing a 10 μm diameter confocal laser spot on the sample. The laser spot size was measured using the image created on an EMCCD camera (Photometrics QuantEM$^{TM}$ 512SC). The laser repetition rate was set to 5 MHz with 11.4 ps resolution to obtain PL decay data. The MoSe$_2$ PL was collected using a MPD visible single photon avalanche diode (Vis-SPAD) *via* 750 nm long pass and 900 nm short pass filters, completely filtering out laser excitation and allowing for collection of MoSe$_2$ PL only. Further precaution was taken to remove any long wave component of the excitation line using a 650 nm short pass filter. All signals were scaled up to 3000 s, which was used in the lowest excitation intensity measurement. Laser power was measured in the excitation line using a Thorlabs S130C photodiode and PM100D power meter. Laser excitation power was regulated using a series of neutral density filters. The instrument response function was measured with a blank glass cover slide as used for the sample.

Decay rates were fitted using a model developed in Origin, which consists of a Gaussian (as the IRF) convoluted with a double exponential decay.

Transistor preparation and characterisation

After exfoliation and transfer onto Si - SiO$_2$ (90nm) isolated monolayer MoSe$_2$ flakes were identified and electrodes with a typical channel length of 4μm were patterned using e-beam lithography and thermal evaporation of Pd:Au (20nm:80nm) followed by lift-off in acetone. Transfer characteristics were measured using a Keithley 4200 SCS connected to a probe station. The global back-gate was swept from negative to positive voltages and the current was measured under a source-drain bias of 5V.

Figure 1.a design

TOC nanocrystal graphics were developed in VESTA software [42] and parsed into ChemDraw3D (Perkin Elmer) for rendering.

**Author contributions**

A.O.A.T prepared samples, performed PL and TRPL measurements, analysed data and wrote manuscript and produced TOC image; J.A.-W and Y.F. performed transistor preparation, electrical characterization and electronic transport data interpretation; N.G. assembled TRPL setup; J.X. provided discussion on surface chemistry; R.P assisted in PL data analysis and; Z.L assisted in TOC image preparation.

**Conflicts of Interest**

There are no conflicts of interest to declare.

**Supporting Information**

Supporting information includes: spectral deconvolution of OA treated $MoSe_2$ PL spectra using Gaussian fits; all time resolved PL signals for pristine and OA treated $MoSe_2$ including bi-exponential decay fits; variation of slow PL decay component with initial carrier concentration and; Raman spectra of pristine and OA treated $MoSe_2$ on glass slide. This material is available free of charge *via* the internet.

**Acknowledgements**

The authors thank the Winton program for physics of sustainability for financial support. We also acknowledge funding from EPSRC Grants EP/L015978/1, EP/L016087/1, EP/P027741/1, EP/M006360/1, and EP/P005152/1. J.A.W. acknowledges the support of his Research Fellowship from the Royal Commission for the Exhibition of 1851, and Royal Society Dorothy Hodgkin Research Fellowship. Z.L. acknowledges funding from the Swedish research council, Vetenskapsrådet 2018-06610. This project has received funding from the European Research Council (ERC) under the European Union's Horizon 2020 research and innovation programme (Grant Agreement 758826). The data underlying this publication are available.

**Figures**

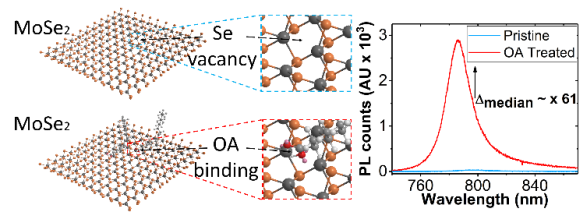

**TOC: PL enhancement in MoSe$_2$ due to OA ligand coordination**

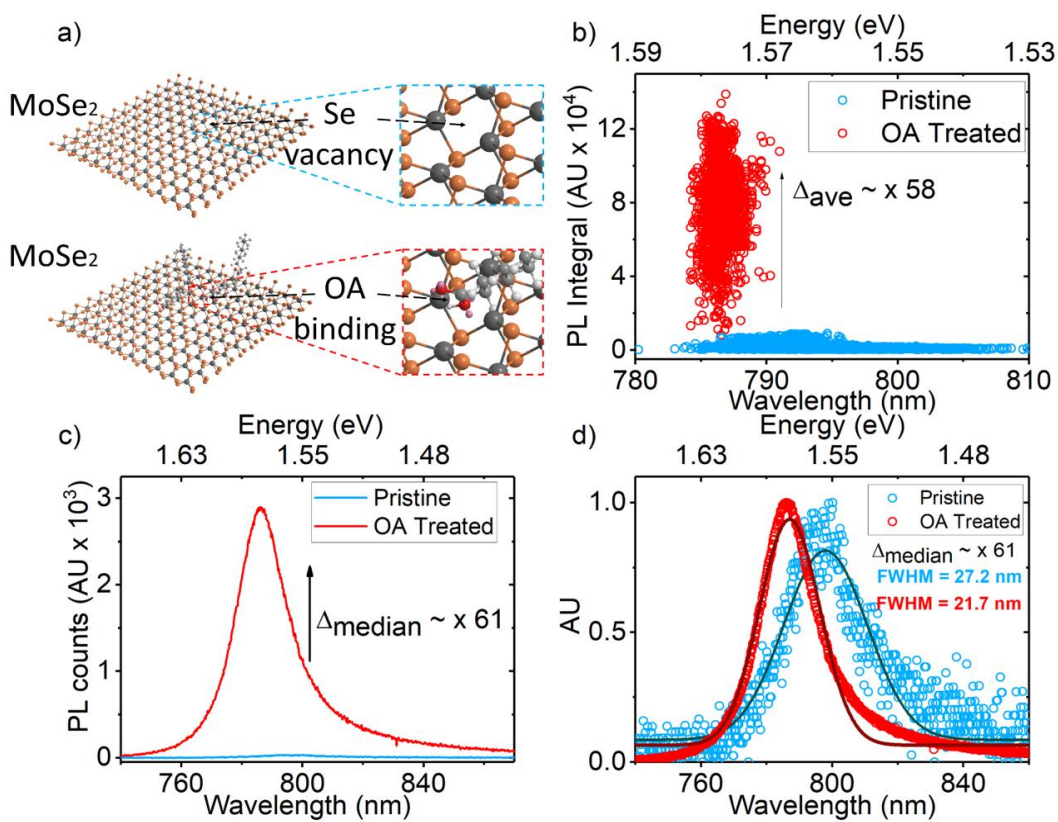

**Figure 1 a-d: a)** Illustration of OA ligand coordination to a selenium (Se) vacancy; **b)** PL enhancement scatter plot showing untreated monolayer PL integrals (blue) and OA treated monolayer PL integrals (red); **c)** Raw PL spectra for points that represent the median PL integrals before (blue) and after OA treatment (red) on an exemplary monolayer; **d)** Normalised median spectra (circles) from Figure 1.c, with single Gaussian peaks (solid lines) fitted to estimate the change in FWHM from untreated (blue) to OA treated (red) case.

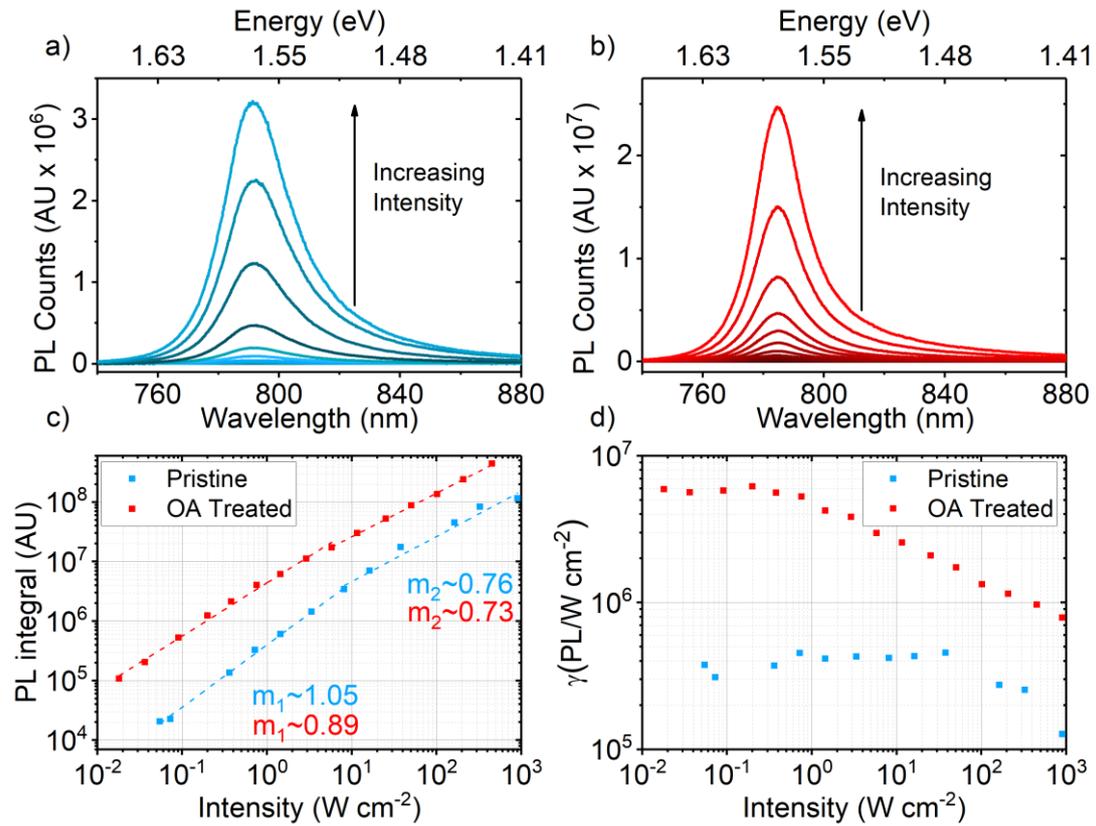

**Figure 2 a-d: a)-b)** Raw PL spectra from excitation intensity series of untreated and OA treated MoSe$_2$ respectively; **c)** Excitation intensity series derived from PL integrals of spectra (a-b) for untreated (blue) and OA treated (red) monolayers; **d)** Ratio of PL integral to excitation intensity *i.e.* relative PLQE (γ) as a function of excitation intensity for untreated (blue) and OA treated (red) monolayers.

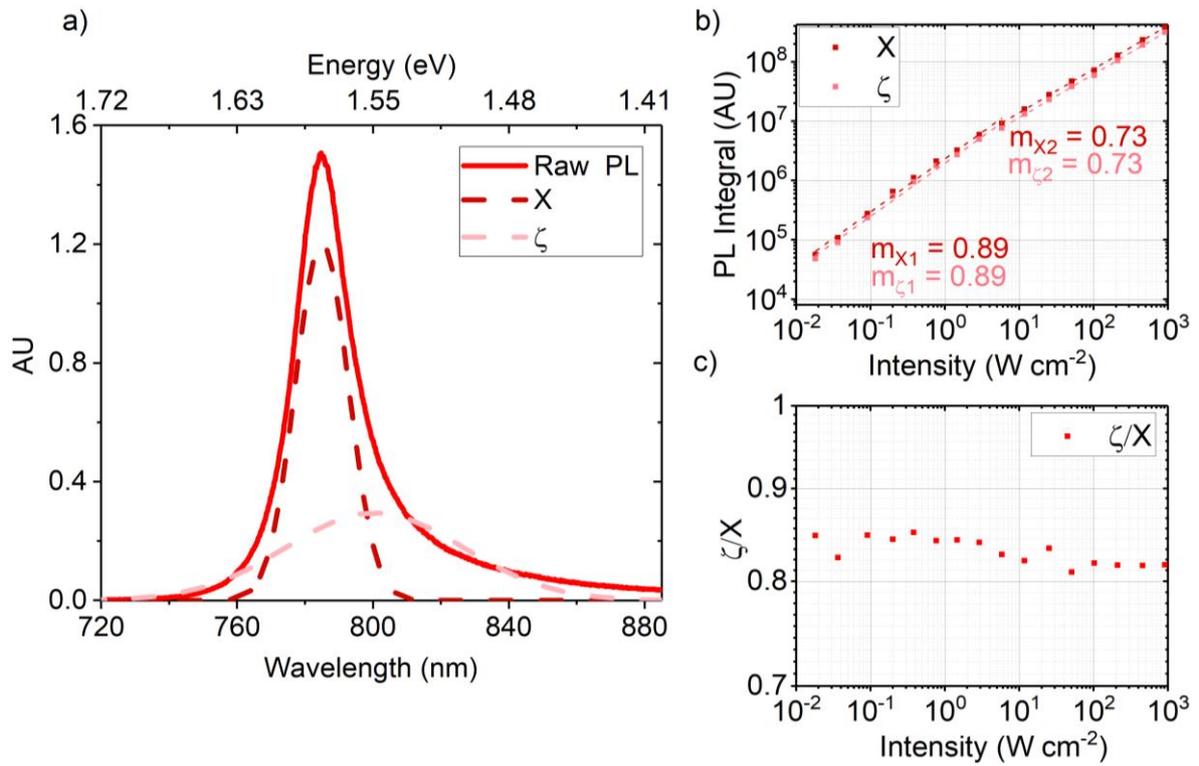

**Figure 3 a-c: a)** Raw PL spectrum of OA treated $MoSe_2$ (red) taken in the high intensity regime (455 W cm$^{-2}$). Dashed maroon and pink Gaussian fits represent the neutral exciton (X) and a low energy species (ζ) respectively; **b)** Excitation intensity series of neutral exciton (X) and low energy species (ζ) showing identical recombination dynamics; **c)** Ratio of ζ to X as a function of excitation intensity. Relatively constant value < 1 implies prevalence of neutral excitons throughout the series.

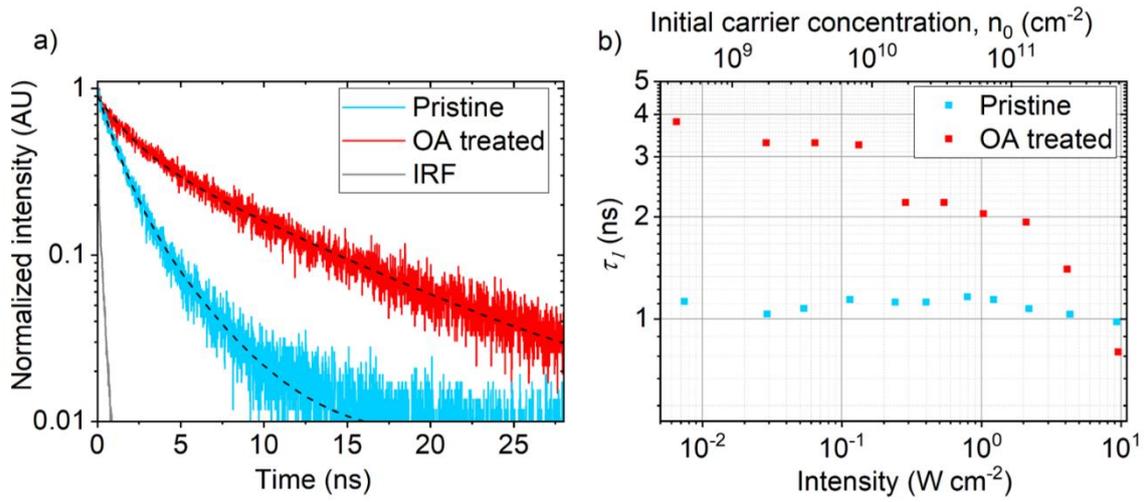

**Figure 4 a-b: a)** Time resolved PL signals of pristine (blue) and OA treated (red) MoSe$_2$ monolayers with bi-exponential decay fits (black dashed lines) measured at comparable pump intensities; 0.054 W cm$^{-2}$ (Pristine sample) and 0.064 W cm$^{-2}$ (OA sample) using 550 nm, 5MHz pulsed excitation.

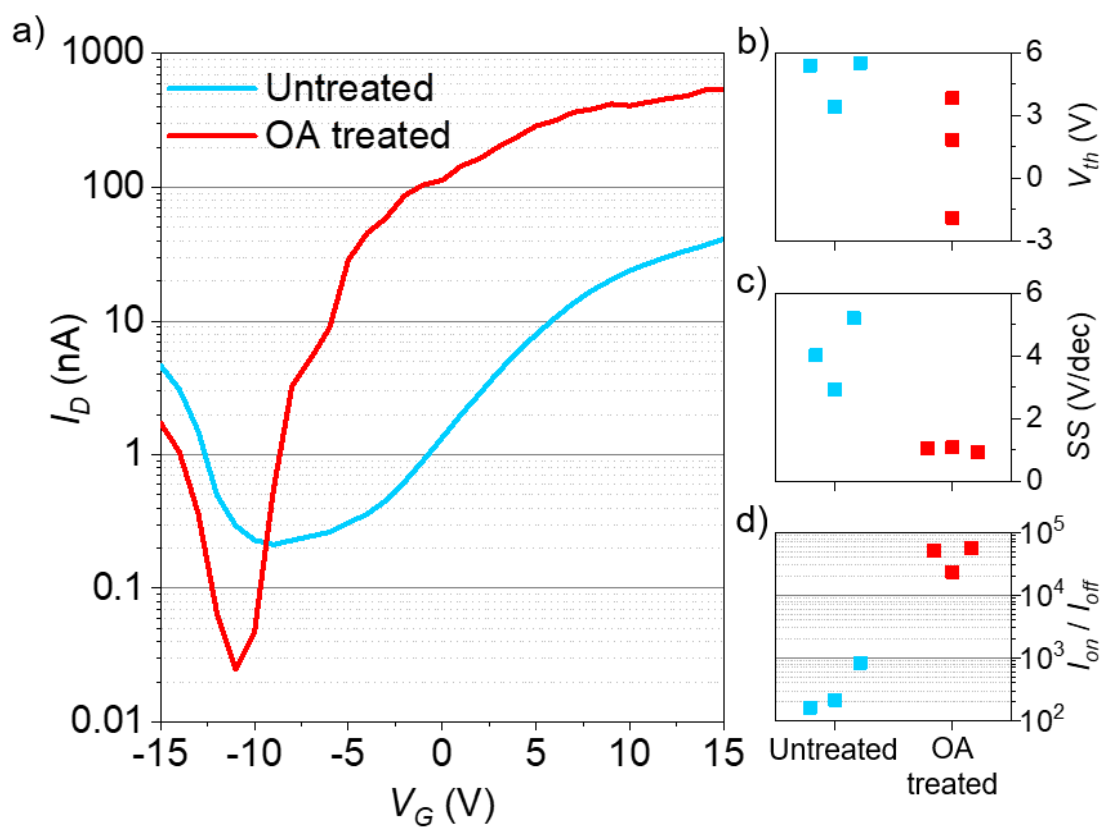

**Figure 5 a-d**: **a)** Transfer characteristics of the same back-gated monolayer MoSe$_2$ field effect transistor before (blue) and after (red) OA treatment. **b)** Threshold voltage, **c)** Subthreshold swing, **d)** On-off current ratio for three MoSe$_2$ transistors before (blue) and after (red) OA treatment.

# Supplementary Information

Arelo O.A Tanoh[1,2], Jack Alexander-Webber[3], Ye Fan[3], Nicholas Gauriot[1], James Xiao[1], Raj Pandya[1], Zhaojun Li[1], Stephan Hofmann[3], Akshay Rao[1]*

[1]Cavendish Laboratory, Cambridge, JJ Thomson Avenue, CB3 0HE, Cambridge, United Kingdom

[2]Cambridge Graphene Centre, University of Cambridge, 9 JJ Thomson Avenue, Cambridge, CB3 0FA, Cambridge, United Kingdom

[3]Department of Engineering, University of Cambridge, JJ Thomson Avenue, CB3 0FA Cambridge, United Kingdom

*E-mail: ar525@cam.ac.uk

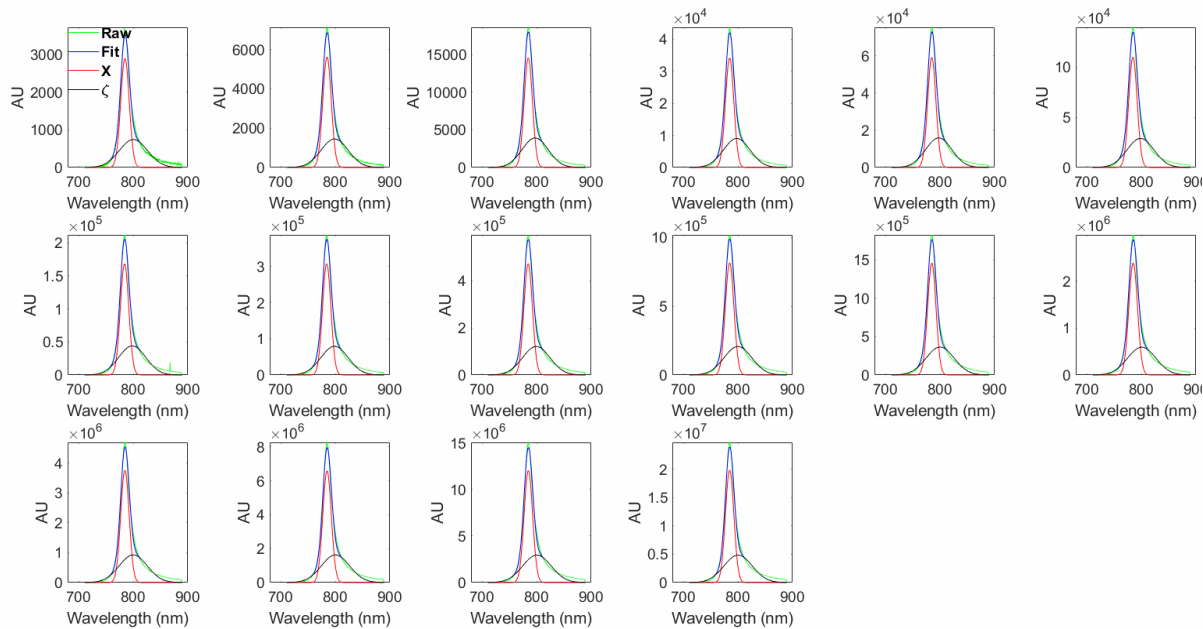

**SI Figure 1**: Gaussian fits of OA treated MoSe$_2$ monolayer PL spectra

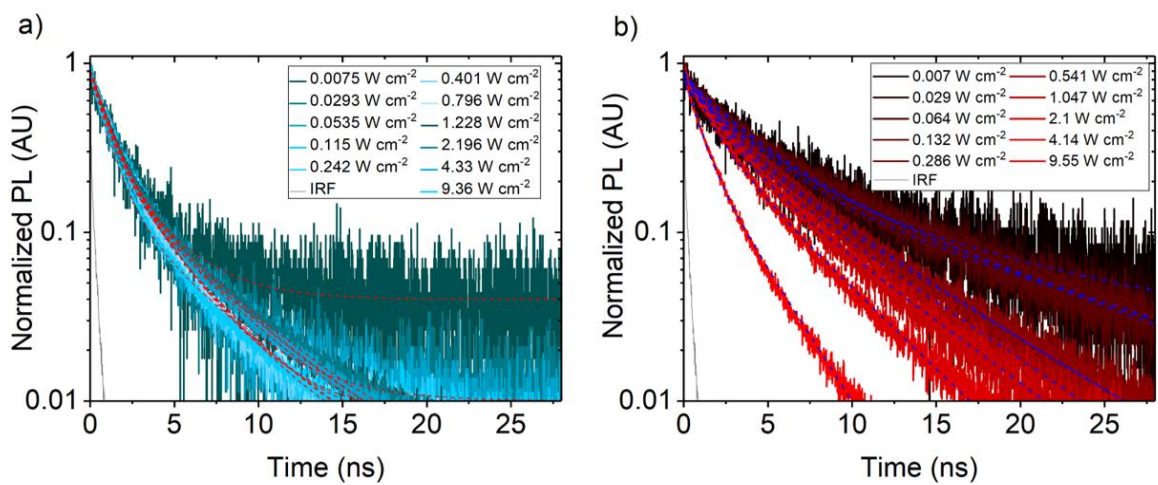

**SI Figure 2 a-b**: Time resolved photoluminescence signals for pristine (blue) and OA treated (red) samples with bi-exponential decay fits (red dashed line in pristine spectra and blue dashed lined in OA treated spectra).

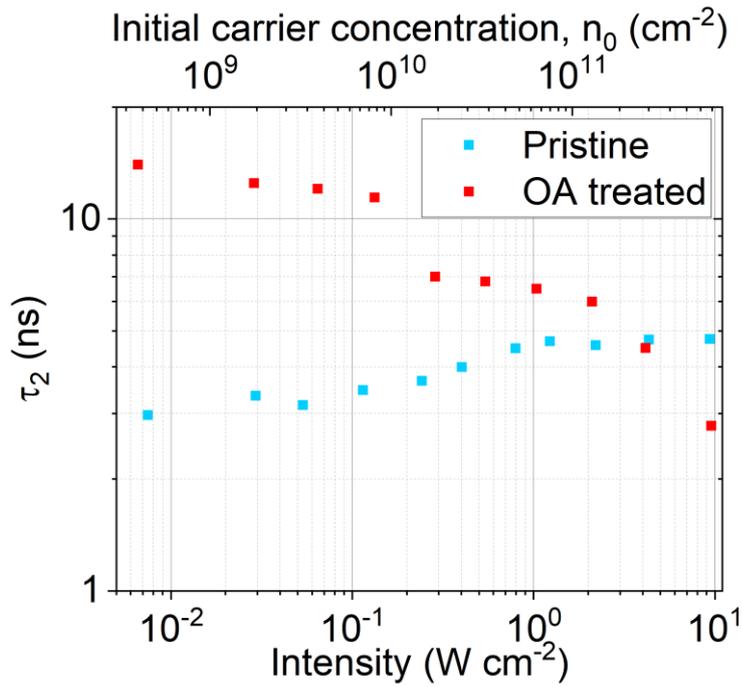

**SI Figure 3:** Variation of slow decay component, $\tau_2$, of pristine (blue) and OA treated (red) $MoSe_2$ time resolved PL signals with initial carrier concentration and pump intensities (W cm$^{-2}$).

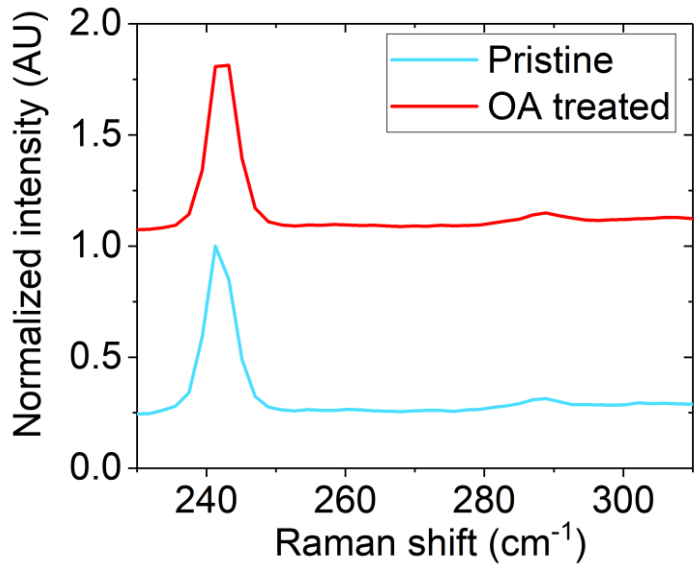

**SI Figure 4**: Raman spectra of pristine (blue) and OA treated $MoSe_2$ monolayers on glass substrate.